%% file: bss-multisite-hal.tex
\documentclass[conference]{IEEEtran}
\IEEEoverridecommandlockouts
\usepackage{cite}
\usepackage{amsmath,amssymb,amsfonts}
\usepackage{graphicx}
\usepackage{textcomp}
\usepackage{xcolor}
\def\BibTeX{{\rm B\kern-.05em{\sc i\kern-.025em b}\kern-.08em
    T\kern-.1667em\lower.7ex\hbox{E}\kern-.125emX}}
\usepackage{color,epsfig,ifthen,twoopt,tabularx,xargs,a4wide,answers}
\usepackage[utf8]{inputenc}
\usepackage[numbers, sort&compress]{natbib}
\usepackage{amsthm}
\usepackage[boxruled,algo2e]{algorithm2e}
\usepackage{enumitem,verbatim}
\usepackage{subcaption}
\usepackage{amsfonts}
\usepackage[bbgreekl]{mathbbol}
\DeclareSymbolFontAlphabet{\mathbbm}{bbold}
\DeclareSymbolFontAlphabet{\mathbb}{AMSb}%

\usepackage[textwidth=2cm,textsize=footnotesize]{todonotes}    

\usepackage[hyphens]{url}
\usepackage{hyperref}
\hypersetup{colorlinks=true,citecolor=blue, urlcolor=blue, linkcolor=black, filecolor=black, breaklinks=true}

\usepackage{cleveref} 
\input{thmdef}

\theoremstyle{remark}
\input{alldefs}

\usepackage{mathrsfs}
\usepackage{stmaryrd}
\begin{document}
\title{Smooth nonnegative tensor factorization for multi-sites electrical load monitoring}

\author{\IEEEauthorblockN{
    Amaury Durand\IEEEauthorrefmark{1}\IEEEauthorrefmark{2},
    François Roueff\IEEEauthorrefmark{1},
    Jean-Marc Jicquel\IEEEauthorrefmark{2},
    and Nicolas Paul\IEEEauthorrefmark{3}} \\
  \IEEEauthorblockA{\IEEEauthorrefmark{1}
    LTCI, Telecom Paris, Institut Polytechnique de Paris. 19 Place Marguerite Perey, 91120 Palaiseau, France.}
  \IEEEauthorblockA{\IEEEauthorrefmark{2} EDF R\&D, TREE, E36, Lab Les Renardieres. Ecuelles, 77818 Moret sur Loing, France.}
  \IEEEauthorblockA{\IEEEauthorrefmark{3}EDF R\&D, PRISME.  6 quai Watier, 78400 Chatou, France. }}

\maketitle 
\begin{abstract}
    The analysis of load curves collected from smart meters is a key step for many energy management tasks ranging from consumption forecasting to customers characterization and load monitoring. In this contribution, we propose a model based on a functional formulation of nonnegative tensor factorization and derive updates for the corresponding optimization problem. We show on the concrete example of multi-sites load curves disaggregation how this formulation is helpful for 1) exhibiting smooth intra-day consumption patterns and 2) taking into account external variables such as the outside temperature. The benefits are demonstrated on simulated and real data by exhibiting a meaningful clustering of the observed sites based on the obtained decomposition.   
\end{abstract}

\section{Introduction}
Since the beginning of consumption data collection, the comparison and disaggregation of load curves have been two popular domains of study. While the former  focuses on clustering rather than extracting meaningful features, the latter consists in decomposing a load curve as a sum of curves each representing a particular consumption pattern. In both cases, it is assumed that a panel of load curves
$\set{X_j(u)}{u \in \cU, \; 1 \leq j \leq J}$
is observed, where $\cU$ is an interval of $\rset$ (e.g. $\cU = [0,24)$) and $j$ is an index (e.g. the day, the customer, the sensor, etc). Numerous clustering methods have been applied to compare such curves (see \cite{CHICCO2012,WEI20181027,Rajabi17} for an overview). On the other hand, load curves disaggregation is usually done in the context of Non Intrusive Load Monitoring (NILM) where the goal is to decompose the globally observed consumption into a sum of the consumption of several devices.  Over the past decade, Blind source separation methods such as Nonnegative matrix factorization (NMF) have gained in popularity in the NILM community (see e.g. \cite{Henriet19NILM} and the references therein). The idea is to decompose the load curve of the $r$-th device as a \emph{signature} curve $a_r(u) \geq 0$ which is modulated across observations by an \emph{activation} $b_{j,r} \geq 0$, that is
$
X_j(u) \approx \sum_{r=1}^R a_r(u) b_{j,r} \; .
$

In a multi-site context, disaggregating the load curves can be a way to extract features to describe the load profiles and cluster the sites. Assume we observe a panel of daily load curves 
$\set{X_{j,n}(u)}{u \in [0,24),  1 \leq j \leq J, \, 1 \leq n \leq N}$
where  $u$ represents the intra-day time, $j$ the observed day and $n$ the site. Generalizing the NMF model of NILM gives the Nonnegative Tensor Factorization (NTF) model, 
$
X_{j,n}(u) \approx \sum_{r=1}^R a_r(u) b_{j,r} c_{n,r}
$,
where $a_r(u) \geq 0$ is the \emph{signature} which is modulated across days by the \emph{day activation}  $b_{j,r} \geq 0$ and across sites by the  \emph{site activation} $c_{n,r} \geq 0$. The signature and activations are refered to as the \emph{factors}. 

Tensor factorization models are very popular in chemometrics and psychometrics but their use for electrical load curves analysis is still recent \cite{Figueiredo2014ExploringTP,Figueiredo15NTFNilm, SANDOVAL2020106431}. Multi-sites load curves disaggregation has been proposed in \cite{SANDOVAL2020106431} using PARAFAC, which is the same as NTF without positivity constraints (see \cite{KoBa09,Cichocki-NTF}). In particular, the authors use the site activations for clustering but do not give any interpretation of the factors nor the clusters. In this paper, we propose several modifications of the NTF model in order to orient the decomposition into more interpretable factors. We also use additional knowledge on the outside temperature and different consumption regimes.  The model is presented in \Cref{sec:our-model} and is validated on simulated and real data in \Cref{sec:expe}. 

The notations used throughout this paper are the following. For all integer $N \geq 1$, $\integersfromoneto{N} = \{1, \cdots, N\}$. For tensor operations, we use the notations of \cite{Cichocki-NTF}, that is $\norm{\cdot}_F$ is the Froebenius norm, $[\cdot]_+$ the positive part, $\oast$ the Hadamard product, $\otimes$ the Kronecker product, $\odot$ the Khatri-Rao product and $\circ$ the tensor product. Note that $\bA^{\oast 2} = \bA \oast \bA$. The unfolding $\bX_{(k)}$ of a tensor $\bX$ is defined in \cite{KoBa09}. 
\section{Proposed model}\label{sec:our-model}
The proposed model is based on NTF with the additional assumption that the consumption $X_{j,n}(u)$ only depends on the day of the year $j$ through two variables : the daily temperature $T_{j,n}$ and the consumption regime $\epsilon_{j,n}$ (e.g. business and non-business days). This leads to the following model 
\begin{equation}\label{eq:spline-ntf}
X_{j,n}(u) \approx \sum_{r=1}^R a_r(u) b_{r}(T_{j,n}) c_{n,r}^{(\epsilon_{j,n})}\; , 
\end{equation}
where $a_r(u) \geq 0, b_r(t) \geq 0, c_{n,r}^{(\epsilon)} \geq 0$. We also assume that the functions $a_r$ and $b_r$ are smooth and that $a_r$ is periodic with a period of $24$ hours (since it represents intra-day behavior). By analogy to the NMF and NTF cases we call the functions $b_r$ the \emph{thermal activations}. 

\subsection{The optimization problem}
Let us consider two grids $(u_1, \cdots, u_I) \subset [0,24) $ and $(t_1, \cdots, t_K) \subset \rset$ which contain all the observed intra-day times and temperatures. Let also $E$ be the number of consumption regimes so that $\epsilon_{j,n} \in \integersfromoneto{E}$ for all $j,n$. Then the factors $a_r, b_r$, $c_r^{(\epsilon)}$, are estimated by solving
\begin{align}
  &\min \left\lbrace F + P  \right\rbrace \nonumber \\
  &\text{such  that for all } r, i,  t,  n, \epsilon, \label{eq:optim-problem}\\
  &\begin{cases}
    a_r(u_i) \geq 0\,, b_r(t_k) \geq 0\, ,  c_{n,r}^{(\epsilon)} \geq 0\, \\
    \int_0^{24} a_r = \int_{t_1}^{t_K} b_r = 1  \,, \\
    a_r(0^+) = a_r(24^-) \\
    a_r'(0^+) = a_r'(24^-)\\
    a_r''(0^+) = a_r''(24^-) \, ,
  \end{cases} \nonumber
\end{align}                                   
and
\begin{align*}
F &= \sum_{i,j,n} \left(X_{j,n}(u_i) - \sum_{r=1}^R a_r(u_i) b_r(T_{j,n}) c_{n,r}^{(\epsilon_{j,n})}\right)^2 \, , \\
P &=  \alpha \sum_{r=1}^R \int_0^{24} (a_r'')^2 + \beta \sum_{r=1}^R \int_{t_1}^{t_K} (b_r'')^2 \; . 
\end{align*}
The scaling constraints prevent the factors from diverging since the error $F$ is not affected by multiplying one of the factors by a constant as soon as the other factors are scaled accordingly. These scaling constraints also make sure that we can compare the site activations.

Finally, the penalizations on the $L^2$-norm of the second derivatives imply that the solutions of Problem \eqref{eq:optim-problem} are necessarily smooth spline functions. Namely, for all $r \in \integersfromoneto{R}$, the function $a_r$ must be a $24$-periodic cubic spline and the function $b_r$ must be a natural cubic spline. Since spline functions are characterized by their sample points, we can reformulate the problem as a weighted NTF. 

\subsection{Formulation as a weighted NTF problem}
Classical results on cubic splines imply that there exist $\bv_1 \in \rset^I$ and $\bQ_1 \in \rset^{I\times I}$ positive definite such that for any 24-period cubic spline $a$ on $[0,24)$, we have $\int_0^{24} a = \bv_1^\top \ba$ and $\int_0^{24} (a'')^2 = \ba^\top \bQ_1 \ba$, with $\ba = [a(u_1), \cdots, a(u_I)]^\top$. Similarly there exist  $\bv_2 \in \rset^K$ and $\bQ_2\in\rset^{K\times K}$ positive definite such that for any cubic spline $b$ on $[t_1, t_K]$, we have $\int_{t_1}^{t_K} b = \bv_2^\top \bb$, and $\int_{t_1}^{t_K} (b'')^2 = \bb^\top \bQ_2 \bb$, with $\bb = [b(t_1), \cdots, b(t_K)]$. 
We now define
$\bA \in \rset_+^{I \times R}$, $\bB \in \rset_+^{K \times R}$
and $\bC \in \rset_+^{EN \times R}$  by $\bA_{i,r} = a_r(u_i), \; \bB_{k,r} = b_r(t_k), \; \bC_{(\epsilon-1) N + n,r} = c_{n,r}^{(\epsilon)}$, 
and $\bW \in \rset^{I \times K \times EN}$, $\bX \in \rset^{I \times K \times EN}$ by
$
\bW_{i,  k,(\epsilon-1) N + n} = \varsqrt{ \sum_{j=1}^J \1_{T_{j,n} = t_k} \1_{\epsilon(j,n) = \epsilon}}
$
and
$$
\bX_{i,k,(\epsilon-1) N + n} = \frac{ \sum_{j=1}^J \1_{T_{j,n} = t_k} \1_{\epsilon_{j,n} = \epsilon} X_{j,n}(u_i)}{\bW_{i,  k,(\epsilon-1) N + n }^2} \;,
$$
with the convention that $0/0 = 0$. 
Then Problem  \eqref{eq:optim-problem} is equivalent to
\begin{equation}\label{eq:optim-problem-samples}
  \begin{split}
    &\min_{\bA \geq 0 , \bB \geq 0, \bC \geq 0}  f_\bW(\bA, \bB, \bC) \\
    & \text{such that for all } r, \bv_1^\top \ba_r = \bv_2^\top \bb_r = 1 \,,
  \end{split}
\end{equation}
where $\ba_r, \bb_r$ and $\bc_r$ are the $r$-th columns of $\bA, \bB$ and $\bC$ respectively, and 
  $$
  f_\bW(\bA, \bB, \bC) = L_\bW(\bA, \bB, \bC) + P(\bA,\bB) \; ,
  $$
  with
  $$
  L_{\bW}(\bA, \bB, \bC) = \norm{\bW \oast \left(\bX - \sum_{r=1}^R \ba_r \circ \bb_r \circ \bc_r\right)}_F^2 \; ,
  $$
  and
  $$
  P(\bA,\bB) =  \alpha \tr(\bA^\top \bQ_1 \bA) +  \beta \tr(\bB^\top \bQ_2 \bB) \; . 
  $$
\subsection{Fast HALS algorithm}
We solve Problem \eqref{eq:optim-problem-samples} using a Fast HALS algorithm. The idea of HALS is to minimize $f_\bW$ alternatively in the columns of $\bA$, $\bB$, $\bC$. The update in $\ba_r$ is obtained by solving
$$
\min_{\substack{\ba_r \geq 0 \\ \bv_1^\top \ba_r = 1}} \norm{\bW \oast (\bX^{(r)} - \ba_r \circ \bb_r \circ \bc_r)^2}_F^2 + \alpha \ba_r^T \bQ_1 \ba_r \, ,
$$
where $\bX^{(r)} = \bX - \sum_{s \neq r} \ba_s \circ \bb_s \circ \bc_s$. This leads to the following update steps
\begin{align*}
  \ba_r &\leftarrow \bM_r^{-1} \left(\bW_{(1)} \oast \bX^{(r)}_{(1)}\right)\left(\bc_r \otimes \bb_r\right) \\
   \ba_r &\leftarrow \frac{[\ba_r]_+}{\bv_1^\top [\ba_r]_+} \; ,
\end{align*}
where $\bM_r := {\rm diag}(\bW_{(1)}^{\oast 2} (\bc_r \otimes \bb_r)^{\oast 2}) + \alpha \bQ_1$. 
The Fast HALS algorithm uses a clever way to write $\bX^{(r)}$ and $\bM_r$ to avoid repeating computations unnecessarily. The updates in $\bb_r$ and $\bc_r$ are obtained similarly. 

\section{Experimental results}\label{sec:expe}
We validate our model on two datasets. The first was extracted from energy demand data simulated for $1$ year and with hourly rate by the Office of Energy Efficiency \& Renewable Energy (EERE)\footnote{Available at \url{https://openei.org/doe-opendata/dataset/commercial-and-residential-hourly-load-profiles-for-all-tmy3-locations-in-the-united-states}.}. We took a total of $775$ sites in California, Arizona, Nevada, Utah, Oregon, Idaho and Washington,  gathering $5$ different building types. The second dataset consists of energy demand collected by EDF from $108$ supermarkets across France over  a period of $1$ year with a sample rate of $10$ minutes. The average external temperature of each day is observed for each site\footnote{EERE's data can be obtained from the TMY3 weather stations using the \texttt{eeweather} Python package.}, EERE's data have one consumption regime and EDF's data have two consumption regimes (closing and opening days). Because of space constraints, we will not discuss the detection of regimes. In order to compare sites of different sizes we scale the observed load curves by the average daily consumption. 

For both NTF and our model, we use the Fast HALS updates and the algorithm is stopped when the relative improvement of the loss reaches $10^{-5}$.  Each factors was initialized by taking the positive part of the singular vectors of the corresponding unfolding of $\bX$ (e.g. $\ba_r = [\bu_r]_+$ where $\bu_r$ is the $r$-th singular vector of $\bX_{(1)}$). Since there is no foolproof method to select the number of components $R$ (see \cite{Timmerman00-ncomp-parafac,Bro03-ncomp-parafac,Ceulemans06-ncomp-parafac} for ad-hoc methods) and since cross-validation is not straightforward with tensor data (see \cite{Bro08-cv,Owen09-cv}), we take $R = 6$ and $\alpha=\beta=3000$ for EERE's data and $\alpha=\beta=50$ for EDF's data. This choice gives a good balance between goodness of fit and interpretation of the factors.
For clustering, we run $K$-means on the sites activations for all regimes. This means that  site $n$ is represented by the feature vector  $(c_{n,1}^{(1)}, \cdots, c_{n,R}^{(1)}, \cdots, c_{n,1}^{(E)}, \cdots, c_{n,R}^{(E)})$ where we recall that $E$ is the number of regimes.  

\subsection{Results on EERE's data}
\begin{figure*}
   \captionsetup[subfigure]{aboveskip=-8pt,belowskip=-5pt}
  \begin{center}
  \begin{subfigure}{0.45\textwidth}
  \begin{center}
    \includegraphics[width=\textwidth]{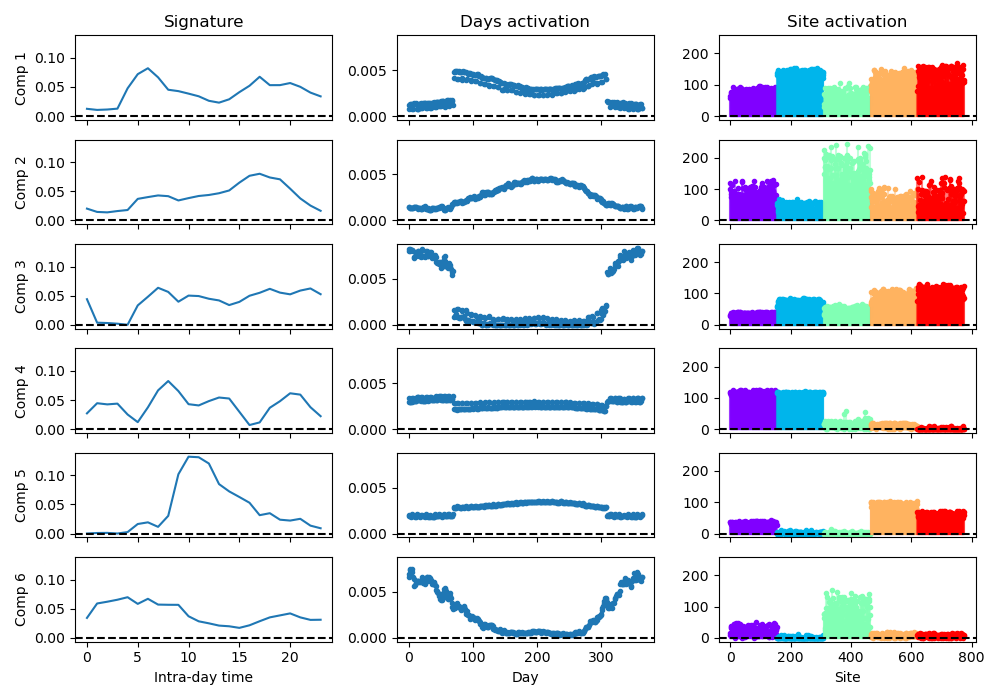}
  \end{center}
  \caption{NTF}
\end{subfigure}
  \begin{subfigure}{0.45\textwidth}
  \begin{center}
    \includegraphics[width=\textwidth]{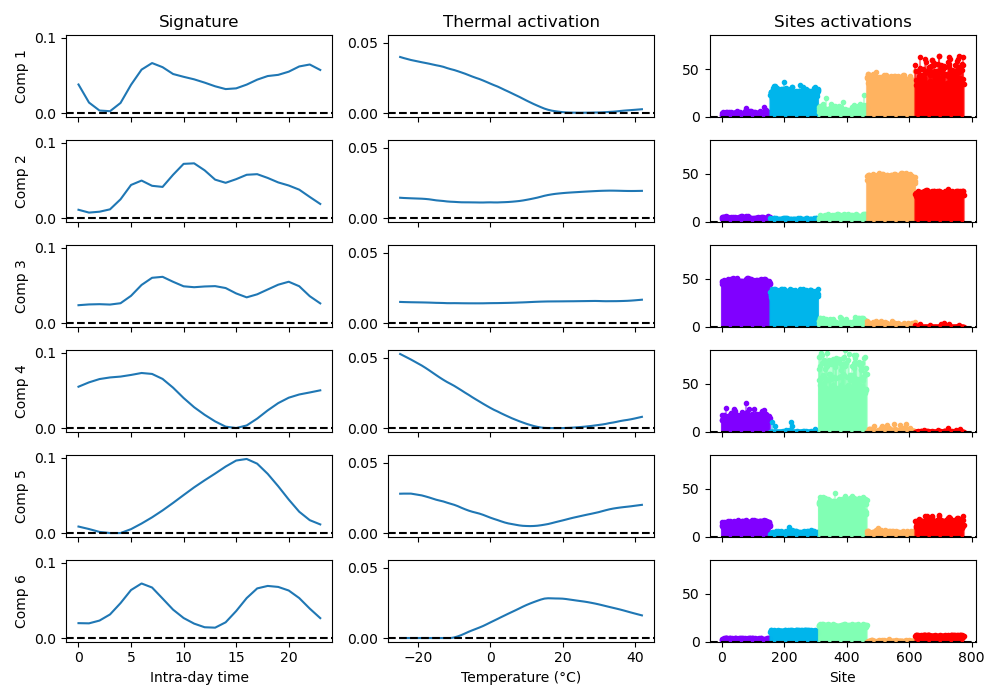}
  \end{center}
  \caption{Our model}
\end{subfigure}
\end{center}
\caption{EERE's Dataset : factors \label{fig:factors-eere}}
\end{figure*}

\begin{figure}
   \captionsetup[subfigure]{aboveskip=-5pt,belowskip=-1pt}
  \begin{center}
  \begin{subfigure}{0.45\textwidth}
  \begin{center}
    \includegraphics[width=\textwidth]{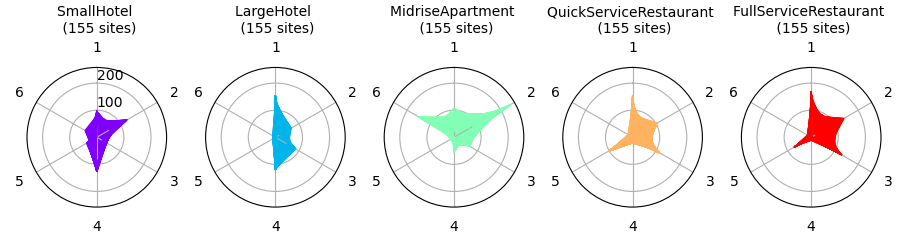}
  \end{center}
  \caption{NTF}
\end{subfigure}
  \begin{subfigure}{0.46\textwidth}
  \begin{center}
    \includegraphics[width=\textwidth]{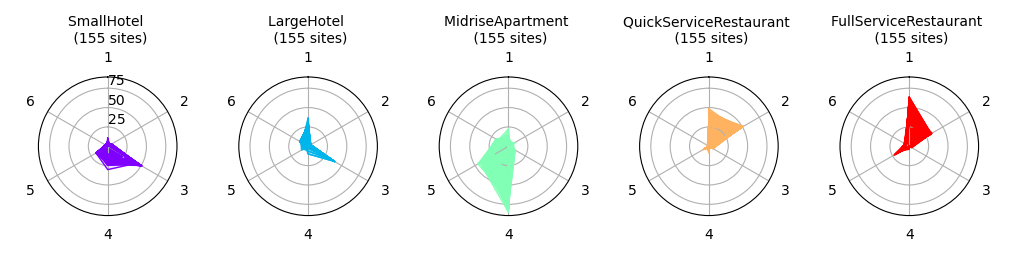}
  \end{center}
  \caption{Our model}
\end{subfigure}
\end{center}
\caption{EERE's Dataset : site activations \label{fig:radar-eere}}
\end{figure}

The factors obtained by NTF and our model are represented in \Cref{fig:factors-eere} where the colors correspond to the building types. The dependence on the temperature is justified by the fact that the day activations of NTF mainly indicate season changes. Moreover, the advantage of smoothing the signatures is that only the most important peaks are kept, which is valuable for interpretation. The radar plots of the site activations presented in \Cref{fig:radar-eere} show a better separation of the building types with our model (especially between the restaurants). To quantify this observation, we ran K-means with $5$ clusters and compared the clusters with the true labels using the adjusted random index. Our model gives a perfect fit with an adjusted random index of $1$ compared to $0.75$ for NTF which mostly fails to separate the two types of restaurants. In our model, both hotels have a high site activation in Component 3 whose signature is typical of a hotel (high for breakfast and dinner and medium for lunch). Small Hotels tend to heat at night (Component 4) while Large Hotels tend to heat during the day (Component 1). Apartments have a high site activation in Component 4 whose signature and thermal activation are typical of heating in residential buildings. Components 5 and 6 also characterize the Apartments and can be interpreted as holidays. Indeed, the thermal activation of Component 5 is high when it is very cold and very hot (winter and summer holidays) where people are more at home in the middle of the day (where the signature is high) while the thermal activation of Component 6 peaks for medium temperatures and its signature presents a typical working day profile. Finally, the signature of Component 2, which peaks before lunch and dinner (and a bit before breakfast), is characteristic of restaurants. The two types of restaurants differ by the importance of lunch (more important for Quick Service Restaurants) and the amount of heating used (Component 1). 

\subsection{Results on EDF's data}
\begin{figure*}
   \captionsetup[subfigure]{aboveskip=-8pt,belowskip=-5pt}
  \begin{center}
  \begin{subfigure}{0.45\textwidth}
  \begin{center}
    \includegraphics[width=\textwidth]{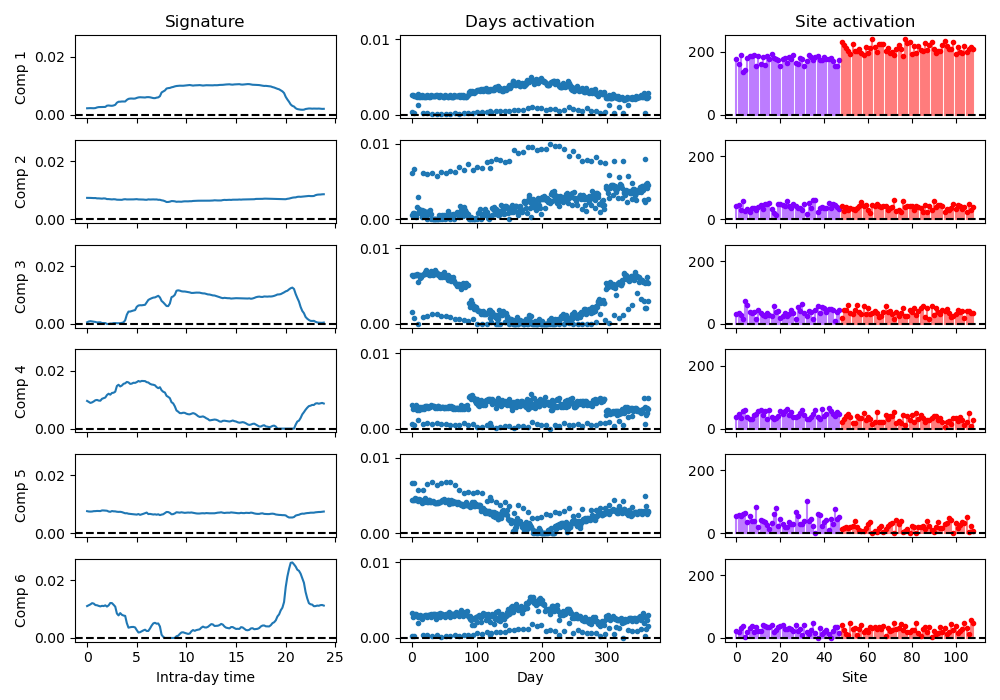}
  \end{center}
  \caption{NTF}
\end{subfigure}
  \begin{subfigure}{0.45\textwidth}
  \begin{center}
    \includegraphics[width=\textwidth]{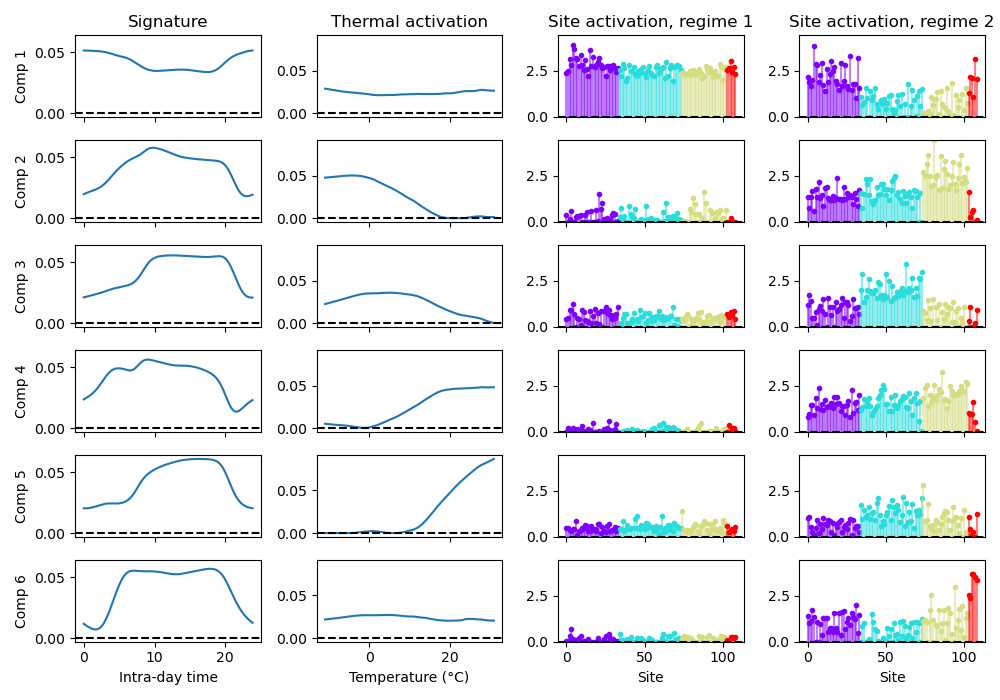}
  \end{center}
  \caption{Our model}
\end{subfigure}
\end{center}
\caption{EDF's Dataset : factors \label{fig:factors-edf}}
\end{figure*}

\begin{figure}
  \begin{center}
    \includegraphics[width=0.4\textwidth]{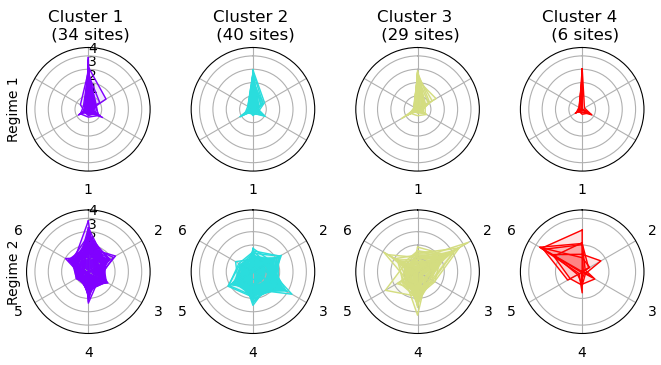}
\end{center}
\caption{EDF's Dataset : site activations of our model \label{fig:radar-edf}}
\end{figure}

The results obtained for EDF's data are presented in \Cref{fig:factors-edf} where the colors correspond to  the clusters obtained by K-means. The number of clusters was selected by the silhouette index (between $2$ and $9$ clusters). The two consumption regimes can be seen in the day activations of NTF which justifies taking them into account in our model. The site activations are not very diverse in  NTF and the two clusters obtained mainly differ from each other by Components 1 and 5 which respectively represent the standard profiles of opening and closing days. On the contrary, our model seems to extract more variable site activations (see also \Cref{fig:radar-edf}) thus exhibiting more clusters. The interpretation of the components is also easier because of smoothness and of the dependence on temperature. For example, the first and last components are independent of the temperature and represent some constant behavior. A high site activation in the  first component implies small difference between opening and closing hours (and therefore is high for regime 1 which represents closing days) and a high site activation in Component 6 implies big difference between opening and closing hours. The other components represent various heating (Components 2 and 3) and cooling (Components 4 and 5) profiles. The signature of Component 2 is characteristic of heater since it peaks in the morning and then  slowly decays because of inertia. The signature and thermal activation of Component 5 indicate that it represents air conditioning  which is activated for high temperatures and in the middle of the day. Finally, the thermal activation of Component 4 indicates that it represents food refrigeration which is activated for lower temperatures than air conditioning and reaches a saturation  level for high temperatures. The clusters can be interpreted as follows.
 Cluster 1 : Small variation between opening and closing hours (Component 1) and small impact of air conditioning (Components 5). 
Cluster 2 : Use a lot of air conditioning  (Components 5) and medium heating profile (Component 3).
 Cluster 3 : High heating  (Component 2) and food refrigeration (Component 4).
 Cluster 4 : Not much impacted by the temperature but the difference between opening and closing hours is high (Components 6).
 
  Giving a physical meaning to the factors and clusters is highly valuable for monitoring or maintenance purpose. For example, sites in Cluster 2 should  reduce their use of air conditioning while for sites in Cluster 3, reducing heating is a priority. 

\section{Conclusion}\label{sec:discussion}
We proposed a model based on a functional formulation of the nonnegative tensor factorization model and an associated optimization algorithm for the disaggregation of multi-sites load curves. By taking into account additional information such as the outside temperature and smoothness of the factors, we showed that this model exhibits more meaningful features and clusters than previously used NTF models.

\end{document}

%% file: thmdef.tex
\crefname{theorem}{theorem}{theorems}
\Crefname{Theorem}{Theorem}{Definitions}

\crefname{definition}{definition}{definitions}
\Crefname{Definition}{Definition}{Definitions}

\crefname{corollary}{corollary}{corollaries}
\Crefname{Corollary}{Corollary}{Corollaries}

\crefname{proposition}{proposition}{propositions}
\Crefname{Proposition}{Proposition}{Propositions}

\crefname{lemma}{lemma}{lemmas}
\Crefname{Lemma}{Lemma}{Lemmas}

\crefname{example}{example}{examples}
\Crefname{example}{Example}{Examples}

\crefname{remark}{remark}{remarks}
\Crefname{remark}{Remark}{Remarks}

\crefname{assumption}{assumption}{assumptions}
\Crefname{assumption}{Assumption}{Assumptions}

\crefname{enumi}{point}{point}
\Crefname{enumi}{Point}{Point}


%% file: alldefs.tex
\newcommand{\integersfromoneto}[1]{[#1]}

\newcommand{\norm}[1]{{
    \left\| #1 \right\|}} 

\newcommandx\cspanarg[2][1=]{\ensuremath{\overline{\mathrm{Span}}^{#1}\left(#2\right)}}

\newcommand{\varsqrt}[1]{\left(#1\right)^{1/2}}

\newcommand{\spec}{{\rm \spec}}

\newcommand{\tnorm}[1]{{\left\vert\kern-0.25ex\left\vert\kern-0.25ex\left\vert #1 
    \right\vert\kern-0.25ex\right\vert\kern-0.25ex\right\vert}} 

\def\rset{\mathbb{R}}

\def\zset{\mathbb{Z}}
\def\nset{\mathbb{N}}

  {\begin{enumerate}}%
  {\end{enumerate}}

\newcommandx{\aslim}[1]{\ensuremath{\stackrel{#1\text{a.s.}}{\longrightarrow}}}  

\newcommand{\1}{\mathbbm{1}}

\def\bA{\mathbf{A}}
\def\ba{\mathbf{a}}
\def\bb{\mathbf{b}}
\def\bB{\mathbf{B}}
\def\bC{\mathbf{C}}
\def\bc{\mathbf{c}}

\def\bM{\mathbf{M}}

\def\bQ{\mathbf{Q}}

\def\bu{\mathbf{u}}

\def\bv{\mathbf{v}}

\def\bW{\mathbf{W}}

\def\bX{\mathbf{X}}

\def\eqsp{\;}

\def\b0{{\bf 0}}
\def\ba{{\bf a}}

\def\bc{{\bf c}}

\def\bu{{\bf u}}

\def\bA{{\bf A}}
\def\bB{{\bf B}}
\def\bC{{\bf C}}

\def\bM{{\bf M}}

\def\bQ{{\bf Q}}

\def\bW{{\bf W}}
\def\bX{{\bf X}}

\def\cU{\mathcal{U}}

\def\tr{\mathrm{Tr}}

\newcommand\algo[1]%
{
    \begin{center} %
    \begin{tabular} {||p{10 cm}l ||}%
               #1 &  \\
    \hline
    \end{tabular}
    \end{center}
}

\newcommand{\chunk}[4][]%
{\ifthenelse{\equal{#1}{}}{\ensuremath{{#2}_{#3:#4}}}{\ensuremath{#2^#1}_{#3:#4}}
}

\def\esp{\mathbb{E}}
\newcommandx\prob[2][1=,2=]{\ensuremath{{\mathbb P}_{#1}^{#2}}}
\newcommand{\PP}[1][]{\ifthenelse{\equal{#1}{}}{\ensuremath{\mathbb{P}}}{\ensuremath{\mathbb{P}\left( #1 \right)}}}
\newcommandx{\PParg}[2][1=]{\PP_{#1}\left(#2\right)}
\newcommand{\PE}[1][]{\ifthenelse{\equal{#1}{}}{\esp}{\ensuremath{{\mathbb E}\left[ #1 \right]}}}
\newcommandx{\PEarg}[2][1=]{\PE_{#1}\left[#2\right]}
\newcommand{\PVar}{\ensuremath{\operatorname{Var}}}
\newcommandx\var[2][1=]{\ensuremath{\PVar_{#1}\left( #2\right)}}
\newcommandx\cvar[3][1=]{\ensuremath{\PVar_{#1}\left( \left. #2 \right| #3 \right)}}

\newcommandx\cov[3][1=]{\ensuremath{\mathrm{Cov}_{#1}\left( #2,#3 \right)}}
\newcommandx\ccov[3][1=]{\ensuremath{\mathrm{Cov}_{#1}\left( \left. #2 \right| #3 \right)}}

\newcommand{\CPP}[3][]
{\ifthenelse{\equal{#1}{}}{\PP\left[\left. #2 \, \right| #3 \right]}{\mathbb{P}_{#1}\left(\left. #2 \, \right | #3 \right)}}
\newcommand{\CPE}[3][]
{\ifthenelse{\equal{#1}{}}{\PE\left[ \left. #2 \right| #3
    \right]}{\mathbb{E}_{#1} \left[ \left. #2 \right| #3 \right]}}
\newcommandx\cprob[4][1=,2=]{\ensuremath{\PP_{#1}^{#2}\left[ \left. #3 \right|
      #4 \right]}}
\newcommandx\HMCP[2][1=]{
\ifthenelse{\equal{#1}{}}{\PP_{#2}}{\PP_{#1,#2}}
}
\newcommandx\HMCE[2][1=]{
\ifthenelse{\equal{#1}{}}{\PE_{#2}}{\PE_{#1,#2}}
}
\newcommandx\HMCEarg[3][1=]{
\ifthenelse{\equal{#1}{}}{\PE_{#2}\left[#3\right]}{\PE_{#1,#2}\left[#3\right]}
}

\def\loikhi2{\mathbf{\chi^2}}

\newcommandx{\proj}[2]{\ensuremath{\operatorname{proj}\left( \left. #1\right|#2\right)}}

\newcommand{\URoot}{\ensuremath{R}}
\newcommand{\UCov}[1][]%
{%
\ifthenelse{\equal{#1}{}}{\URoot \URoot^t}{\URoot_{#1} \URoot^t_{#1}}%
}

\newcommand{\VRoot}{\ensuremath{S}}
\newcommand{\VCov}[1][]%
{%
\ifthenelse{\equal{#1}{}}{\VRoot \VRoot^t}{\VRoot_{#1} \VRoot^t_{#1}}%
}

\newcommand{\LDX}[2]{\ensuremath{L}}

\newcommand{\postdx}[3][]%
{%
\ifthenelse{\equal{#1}{}}{\ensuremath{\psi_{#2|#3}}}{\ensuremath{\psi_{#1,#2|#3}}}%
}

\newcommand{\epostdx}[3][]%
{%
\ifthenelse{\equal{#1}{}}{\ensuremath{\hat{\psi}_{#2|#3}}}{\ensuremath{\hat{\psi}_{#1,#2|#3}}}%
}

\newcommandx{\predx}[3][1=\bX]{#1_{#2|#3}}   
\newcommand{\predpx}[3][]%
{%
\ifthenelse{\equal{#1}{}}{\ensuremath{\varphi_{#2|#3}}}{\ensuremath{\varphi_{#1,#2|#3}}}%
}
\newcommandx\cesp[4][1=,2=]{\ensuremath{{\mathbb E}_{#1}^{#2}\left[ \left. #3 \right| #4 \right]}}

\newcommand{\filt}[2][]%
{%
\ifthenelse{\equal{#1}{}}{\ensuremath{\phi_{#2}}}{\ensuremath{\phi_{#1,#2}}}%
}
\newcommand{\pred}[3][]%
{%
\ifthenelse{\equal{#1}{}}{\ensuremath{\phi_{#2|#3}}}{\ensuremath{\phi_{#1,#2|#3}}}%
}
\newcommand{\post}[3][]%
{%
\ifthenelse{\equal{#1}{}}{\ensuremath{\phi_{#2|#3}}}{\ensuremath{\phi_{#1,#2|#3}}}%
}
\newcommand{\logl}[2][]%
{%
\ifthenelse{\equal{#1}{}}{\ensuremath{\ell_{#2}}}{\ensuremath{\ell_{#1,#2}}}%
}
\newcommand{\lhood}[2][]%
{%
\ifthenelse{\equal{#1}{}}{\ensuremath{\mathrm{L}_{#2}}}{\ensuremath{\mathrm{L}_{#1,#2}}}%
}
\newcommand{\cc}[2][]%
{%
\ifthenelse{\equal{#1}{}}{\ensuremath{c_{#2}}}{\ensuremath{c_{#1,#2}}}%
}
\newcommand{\forvar}[2][]%
{%
\ifthenelse{\equal{#1}{}}{\ensuremath{\alpha_{#2}}}{\ensuremath{\alpha_{#1,#2}}}%
}
\newcommand{\nforvar}[2][]%
{%
\ifthenelse{\equal{#1}{}}{\ensuremath{\bar{\alpha}_{#2}}}{\ensuremath{\bar{\alpha}_{#1,#2}}}%
}

\newcommand{\BK}[2][]%
{%
\ifthenelse{\equal{#1}{}}{\ensuremath{\mathrm{\mathrm{B}}_{#2}}}{\ensuremath{\mathrm{B}_{#1,#2}}}%
}
\newcommand{\filtfunc}[2][]%
{%
\ifthenelse{\equal{#1}{}}{\ensuremath{\tau_{#2}}}{\ensuremath{\tau_{#1,#2}}}%
}

\newcommand{\filtmean}[2][]
{\ifthenelse{\equal{#1}{}}{{\ensuremath{\hat{X}_{#2|#2}}}}{\ensuremath{\hat{X}_{#1,#2|#2}}}
}
\newcommand{\filtcov}[2][]
{\ifthenelse{\equal{#1}{}}{\ensuremath{\Sigma_{#2|#2}}}{\ensuremath{\Sigma_{#1,#2|#2}}}}
\newcommand{\postmean}[3][]
{\ifthenelse{\equal{#1}{}}{\ensuremath{\hat{X}_{#2|#3}}}{\ensuremath{\hat{X}_{#1,#2|#3}}}
}
\newcommand{\postcov}[3][]
{\ifthenelse{\equal{#1}{}}{\ensuremath{\Sigma_{#2|#3}}}{\ensuremath{\Sigma_{#1,#2|#3}}}}

\newcommandx{\QEM}[4][1=,4=]{\ensuremath{\mathcal{Q}_{#1}(#4;#2 \, ; #3)}}

\newcommandx\sequence[3][2=t,3=\zset]{\ensuremath{\left(#1_{#2}\right)_{#2 \in #3 }}}
\newcommandx\dsequence[4][3=t,4=\zset]{\ensuremath{\left( (#1_{#3}, #2_{#3})\right)_{#3 \in #4}}}
\newcommandx{\sequencen}[2][2=n\in\nset]{\ensuremath{\left(#1\right)_{#2}}}

\def\bpm{\left[\begin{matrix}}
\def\epm{\end{matrix}\right]}
\def\bma{\begin{matrix}}
\def\ema{\end{matrix}}

\newcommand{\be}{\begin{equation}}     
\newcommand{\ee}{\end{equation}}        

\newcommandx\lnorm[3][1=]{\left\lVert #2 \right\rVert^{#1}_{#3}}

\newcommandx\supnorm[2][1=]{| #2 |^{#1}_\infty}

\newcommandx\ball[3][1=]{\mathrm{B}_{#1} (#2,#3)}

\newcommandx{\prohosym}[1][1=]{{\boldsymbol\rho}_{#1}}
\newcommandx{\proho}[3][1=]{\prohosym{#1}\left(#2,#3\right)}

\newcommandx{\pp}[1][1=\mu]{\ensuremath{#1\-\mathrm{a.e.}}}

\renewcommand{\-}{\mbox{-}}

\newcommandx{\as}[1][1=\PP]{\ensuremath{#1\-\mathrm{a.s.}}}

\newcommandx{\oscnorm}[3][1=,3=]{\operatorname{osc}^{#1}_{#3}\left(#2\right)}

\newcommandx{\tvdist}[3][1=]{\ensuremath{d^{#1}_{\mathrm{TV}}}(#2,#3)}

\newcommandx{\VnormFunc}[3][1=]{\ensuremath{\left|#2\right|_{\mathrm{#3}}^{#1}}}

\newcommand{\continuousfunctionset}[1]{\mathrm{C}_b(#1)}

\newcommand{\lipschitzfunctionset}[1]{\mathrm{Lip}(#1)}
\newcommand{\boundedlipschitzfunctionset}[1]{\mathrm{Lip}_b(#1)}

\newcommandx\functionsetarg[2][1=]{
\ifthenelse{\equal{#1}{c}}{\continuousfunctionset{\mathsf{#2}}}
{\ifthenelse{\equal{#1}{bc}}{\mathrm{C}_b(#2)}
{\ifthenelse{\equal{#1}{u}}{\mathrm{U}(#2)}
{\ifthenelse{\equal{#1}{bu}}{\mathrm{U}_b(#2)}
{\ifthenelse{\equal{#1}{l}}{\lipschitzfunctionset{#2}}
{\ifthenelse{\equal{#1}{bl}}{\boundedlipschitzfunctionset{#2}}
{\mathbb{F}_{#1}(#2)}
}}}}}}

\newcommandx\functionsetspec[1][1=]{
\ifthenelse{\equal{#1}{c}}{\mathrm{C}}
{\ifthenelse{\equal{#1}{bc}}{\mathrm{C}_b}
{\ifthenelse{\equal{#1}{u}}{\mathrm{U}}
{\ifthenelse{\equal{#1}{bu}}{\mathrm{U}_b}
{\ifthenelse{\equal{#1}{l}}{\mathrm{Lip}}
{\ifthenelse{\equal{#1}{bl}}{\mathrm{Lip}_b}
{\mathbb{F}_{#1}}
}}}}}}

\newcommandx{\taboo}[3][1=,3=]{\left(\leftidx{_#1}{#2}{}\right){^{#3}}}

\newcommandx\vectornorm[2][1=]{\left| #2 \right|^{#1}}  

\newcommand{\ensemble}[2]{\left\{#1\,:\eqsp #2\right\}}
\newcommand{\set}[2]{\ensemble{#1}{#2}}

\newcommandx{\plim}[1]{\ensuremath{\stackrel{#1\-\text{prob}}{\longrightarrow}}}
\newcommandx{\dlim}[1]{\ensuremath{\stackrel{#1}{\Longrightarrow}}}

\newcommandx\measureset[3][1=\mathrm{s},3=]{\mathbb{M}^{#3}_{#1}(#2)}
\newcommandx\measuresetmetric[2][1=1]{\mathbb{M}_{#1}(\mathcal{B}(\mathsf{#2}))}  
\newcommandx\measuresetspec[1][1=\mathrm{s}]{\mathbb{M}_{#1}}

\newcommandx\canonicalkernel[1][1=P]{\mathbb{K}_{#1}}